\documentclass[superscriptaddress, twocolumn, amsmath, amssymb, aps,prl, letterdraft, notitlepage,longbibliography]{revtex4-1}
\usepackage{graphicx,graphics,epsfig,subfigure,times,bm,bbm,amssymb,amsmath,amsfonts,amsthm,mathrsfs,MnSymbol}
\usepackage[matrix,frame,arrow]{xypic}
\usepackage[normalem]{ulem}
\usepackage{slashed}
\usepackage{color}
\usepackage[usenames,dvipsnames,svgnames,table]{xcolor}
\usepackage[english]{babel}
\usepackage{verbatim}
\usepackage{tabularx}
\definecolor{darkblue}{rgb}{0.0,0.0,0.3}
\usepackage[colorlinks=true,
            linkcolor=red,
            urlcolor= darkblue,
            citecolor=blue]{hyperref}

\newcommand{\bea}{\begin{eqnarray}}
\newcommand{\eea}{\end{eqnarray}}

\begin{document}
\title{Sharp negative differential resistance from vibrational mode softening in molecular junctions}  

\author{Junjie Liu}
\address{Department of Chemistry and Centre for Quantum Information and Quantum Control,
University of Toronto, 80 Saint George St., Toronto, Ontario, M5S 3H6, Canada}
\author{Dvira Segal}
\address{Department of Chemistry and Centre for Quantum Information and Quantum Control,
University of Toronto, 80 Saint George St., Toronto, Ontario, M5S 3H6, Canada}
\address{Department of Physics, 60 Saint George St., University of Toronto, Toronto, Ontario, Canada M5S 1A7}

\begin{abstract}
We unravel the critical role of vibrational mode softening  
in single-molecule electronic devices at high bias. 
Our theoretical analysis is carried out with a minimal model for molecular junctions, with mode softening arising due to quadratic electron-vibration couplings, and by developing a mean-field approach. We discover that the negative sign of  the quadratic electron-vibration coupling coefficient 
can realize at high voltage a sharp negative differential resistance (NDR) effect with a large peak-to-valley ratio. Calculated current-voltage characteristics, obtained based on {\it ab initio} parameters
for a nitro-substituted oligo(phenylene ethynylene) junction, agree very well with measurements.
Our results establish that vibrational mode softening is a crucial effect at high voltage, 
underlying NDR, a substantial diode effect, and the 
breakdown of current-carrying molecular junctions.
\end{abstract}

\date{\today}

\maketitle

{\it Introduction.}--
Can the field of molecular electronics accomplish its potential and bring to fruition single-molecule devices 
\cite{CuevasB,Aradhya.13.NN,Xiang.16.CR,Evers.19.NULL,Gehring.19.NRP}?
Historically, with the objective to complement and even replace traditional semiconductor devices,
research was largely focused on (relatively) high-voltage applications (up to 4V) \cite{Reed.97.S}.
This includes, for example, diodes based on molecular-junctions (MJs)  \cite{Aviram,Elbing.05.PNAS,Diez.09.NC,Batra.13.NL,Perrin.15.JPCC,Capozzi.15.NN}, transistors \cite{Kubatkin.03.N,Perrin.15.CSR}, and switches \cite{Molen.10.JP,Ke.19.I}. 
However, with the challenge to stabilize performance, reproduce results, and interpret the current-voltage (IV) characteristics at high bias, recent 
research had largely emphasized the more basic transport quantity, that is the linear response electrical conductance
(as well as the thermopower, another linear transport coefficient). Indeed, at low applied voltage many single-molecule transport experiments were successfully explained using the noninteracting coherent electron transport picture, with the Landauer formula backed by {\it ab initio} parametrization of the junction, a handful of examples include \cite{Latha3,TaoR,Aradhya.13.NN,Oren1,Latha1,Latha2,Capozzi.16.NL,CuevasB}.

Among many intriguing  {\it high-voltage} molecular functionalities, arguably the most attractive one is the negative differential resistance (NDR) effect, which had played a seminal role in the field of semiconductor electronics. NDR is a nonlinear effect: in a certain region of the IV curve, increasing the applied voltage results in the suppression of charge current. Stimulated by promising electronic applications that an NDR effect can offer, it has been extensively investigated in a variety of single MJs
\cite{Gaudioso.00.PRL,Xiao.05.JACS,Choi.06.PRL,Mentovich.08.S,Kang.10.APL,Zhou.13.N,Perrin.14.NN,Capozzi.15.NN,Capozzi.16.NL,Perez17,Kuang.18.JACS,Fung.19.NL} and self-assembled mononlayers \cite{Chen.S.99,Xue.99.PRB,Chen.00.APL,Le.03.APL,Nij1}.
To elucidate underlying mechanisms, a great body of theoretical analysis \cite{Seminario.00.JACS,Boese.01.EPL,Karzazi.01.JACS,Emberly.01.PRB,Karzazi.03.N,Galperin.05.NL,Zazunov.06.PRB,Muralidharan.07.PRB,Yeganeh.07.JACS,Cardamone.08.PRB,Galperin.S.08,Galperin.08.JP,Pati.08.PRL,Han.10.PRB,Kaasbjerg.11.PRB,Hartle.11.PRB,Dzhioev.12.PRB,Perfetto.13.PRB,Migliore.13.JACS,Migliore.11.ACSN,Dubi.13.JCP,Xu.15.JP} with {\it ab initio} simulations, or model-system calculations have been carried out. Nevertheless, physical processes behind sharp NDR behaviors 
\cite{Xiao.05.JACS,Fung.19.NL} and vibrational instability remain elusive. 

Developing functional molecular electronic devices hinges on fundamental understanding of key interactions in the system, 
most importantly, the coupling of conducting electrons to intra- and inter-molecular vibrational modes \cite{CuevasB,Galperin.07.JP}.
At low voltage, signatures of electron-vibration couplings are typically modest, and oftentimes can be treated in a perturbative manner  \cite{CuevasB,Galperin.07.JP}. In contrast, at high voltage vibrational excitations become substantial, leading to  conformational change, heating, structural instability, and eventual junctions' rupture.

So far, theoretical analysis of this problem were almost exclusively focused on the {\it linear} 
electron-vibration coupling (LEVC) model \cite{Thomas07,Galperin.07.JP}, assuming small displacements from equilibrium. 
While theoretical studies of higher-order EVCs effects in MJs are scarce, recent experiments have highlighted the significance of {\it quadratic} electron-vibration couplings (QEVC) in molecules. A prominent outcome of QEVC is current-induced vibrational mode softening \cite{Kaasbjerg.13.PRB}, as revealed by Raman spectroscopy measurements on current-carrying MJs \cite{Ward.11.NN,Mirjani.12.JPCC,Li.14.PNAS}. This effect is associated with the breakdown of MJs under large bias, thereby representing a generic feature of MJs. Another novel example shows in temperature-dependent emission spectra measurements of single organic molecules \cite{Clear.20.PRL}. 
It is highly desirable to obtain a better understanding of whether, and how QEVC 
 impacts the functionality of molecular devices.
Since at high voltage the nuclei explore configurations further and further away from equilibrium,
 it is conspicuous that one cannot ignore then high-order EVCs, though this omission has been the norm in the field.

In this Letter, we discover that vibrational mode softening, an outcome of QEVC \cite{Kaasbjerg.13.PRB}, can realize a sharp (abrupt) NDR effect in molecular devices. 
As such, we clear up two puzzles:  (i) We provide a mechanism for an abrupt NDR effect in MJs, as observed in experiments from the early days of molecular electronics \cite{Chen.S.99,Xue.99.PRB,Chen.00.APL,Gaudioso.00.PRL,Le.03.APL,Xiao.05.JACS,Choi.06.PRL,Mentovich.08.S,Kang.10.APL,Zhou.13.N,Perrin.14.NN,Capozzi.15.NN,Capozzi.16.NL,Kuang.18.JACS}. 
(ii) We show that popular LEVC models are insufficient to explain transport behavior
far from equilibrium, which could explain 
discrepancies between experiments and modelling, see e.g. \cite{Galperin.05.NL,Galperin.S.08,Fung.19.NL}. 

To make the physical picture transparent, we carry out a microscopic analysis within a minimal quantum transport model of a single 
spin-degenerate electronic level. Both the LEVC and QEVC are taken into account with the latter accounting for mode softening in MJs \cite{Kaasbjerg.13.PRB}. Motivated by the separation of timescales in MJs displaying nonlinear transport behavior \cite{Galperin.05.NL,Migliore.11.ACSN,Migliore.13.JACS,Fung.19.NL}, we put forward an effective description for the coupled electron-vibration dynamics with a renormalized molecular Hamiltonian, which is valid in both adiabatic (fast electrons) and nonadiabatic (fast vibrations) regimes. 
We show that the sharp NDR results from a 
prominent feature of QEVC 
 in electron-conducting MJs: the corresponding coupling coefficients are generally negative \cite{Ward.11.NN,Kaasbjerg.13.PRB,Li.14.PNAS}. 
 The calculated current-voltage characteristics, obtained in a self-consistent manner based on {\it ab initio} parameters, convincingly reproduces the observed abrupt NDR behavior in electron-conducting nitro-substituted oligo(phenylene ethynylene) (OPE-NO$_2$) single-molecule break junctions as examined in Ref. \cite{Xiao.05.JACS}. Our study thereby establishes a generic mechanism underlying NDR behavior in MJs. More generally, it opens up a route to modelling high-voltage nonlinear single-molecule devices, which has been the original mission of molecular electronics.  

{\it Model and physical mechanism.}--We use a minimum description for the MJ to capture the impact of mode softening under applied bias. 
Our model includes a single spin-degenerate electronic level representing the lowest unoccupied molecular orbital (LUMO) with a `bare' energy $\epsilon_0$ and electronic annihilation operators $d_{\sigma}$ with spin index $\sigma=\uparrow,\downarrow$. This level (molecule) is sandwiched between two metallic leads with annihilation operators $c_{kv\sigma}$ ($v=L,R$). The total Hamiltonian $H=H_M+H_E$ contains the following  parts. The molecular Hamiltonian $H_M$ describes the LUMO electronic level,  a local prominent molecular vibration with the annihilation operator $b$ and a `bare' frequency $\omega_b$, and EVCs up to {\it quadratic} order in the vibrational displacement (setting $e=1$, $\hbar=1$, $k_B=1$ and Fermi energy $\epsilon_F=0$ hereafter),
\begin{equation}\label{eq:hm}
H_M~=~\left[\epsilon_0+\lambda \omega_b(b^{\dagger}+b)+\eta\omega_b(b^{\dagger}+b)^2\right]\sum_{\sigma}n_{\sigma}+\omega_bb^{\dagger}b.
\end{equation} 
Here, $n_{\sigma}\equiv d^{\dagger}_{\sigma}d_{\sigma}$, $\lambda$ and $\eta$ denote the LEVC and QEVC coefficients, respectively. We maintain the spin index to account for spin degeneracy. We do not include explicit Coulomb interactions,
but limit the level occupancy, as break-junction experiments are often performed in the sequential tunneling regime \cite{CuevasB,Gehring.19.NRP}. 
We note that for electron-conducting MJs, $\eta$ is typically negative, corresponding to mode softening upon charging \cite{Ward.11.NN,Kaasbjerg.13.PRB,Li.14.PNAS}. While here we include only a single vibration mode, we emphasize that EVC coefficients are determined in such a way that the primary mode can in fact 
represent a collective effect (for instance, the total reorganization energy) from many active vibrations.
One may also incorporate into the model the coupling of the primary mode to a secondary thermal bath, as we discuss in Ref. \cite{SM.20.NULL}. However, the secondary bath does not play a significant role in our NDR effect at weak coupling \cite{Galperin.05.NL,Galperin.07.JP}. 

The second part of the Hamiltonian includes the two metallic leads and the electron tunneling coupling, $H_E=\sum_{kv\sigma}\epsilon_{kv}c_{kv\sigma}^{\dagger}c_{kv\sigma}+\sum_{kv\sigma}t_{kv}(c_{kv\sigma}^{\dagger}d_{\sigma}+d^{\dagger}_{\sigma}c_{kv\sigma})$. Here, $c_{kv\sigma}$ annihilates an electron with a spin index $\sigma$ and energy $\epsilon_{kv}$ in the $v$-lead, $t_{kv}$ denotes the spin-independent tunneling energy. We introduce spectral densities for the metallic leads as $\Gamma_v(\epsilon)=\pi\sum_{k}t^2_{kv}\delta(\epsilon-\epsilon_{kv})$ and consider the wideband limit, 
$\Gamma_v(\epsilon)=\Gamma_v$ throughout the study  \cite{Wingreen.89.PRB}.

Motivated by the time scale separation in MJs exhibiting nonlinear transport behavior \cite{Galperin.05.NL,Migliore.11.ACSN,Migliore.13.JACS,Fung.19.NL}, we split operators ($A$) into their steady state expectation values ($\bar{A}\equiv\langle A\rangle$) plus fluctuation terms ($\delta A$),
\begin{equation}\label{eq:splitting}
n_{\sigma}~=~\bar{n}_{\sigma}+\delta n_{\sigma},~~b~=~\bar{b}+\delta b.
\end{equation}
In the Heisenberg picture, only the fluctuation term carries the time dependence of the original operator. Applying Eq. (\ref{eq:splitting}) to EVC terms in Eq. (\ref{eq:hm}) and keeping terms containing either $\delta n_{\sigma}$ or $\delta b$, we get $\left[\lambda \omega_b(b^{\dagger}+b)+\eta\omega_b(b^{\dagger}+b)^2\right]\sum_{\sigma}\bar{n}_{\sigma}+\left[\lambda \omega_b\sqrt{2\omega_b}\bar{x}_b+2\eta\omega_b^2\bar{x}_b^2\right]\sum_{\sigma}n_{\sigma}$ where $\bar{x}_b\equiv(\bar{b}^{\ast}+\bar{b})/\sqrt{2\omega_b}$. Here, we added constant terms $(\lambda\omega_b\sqrt{2\omega_b}\bar{x}_b+2\eta\omega_b^2\bar{x}_b^2)\sum_{\sigma}\bar{n}_{\sigma}$, which do not modify the dynamics, and neglected product terms proportional to $\delta x_b\delta n_{\sigma}$ and $\delta x_b^2\delta n_{\sigma}$. 
 Since electrons and vibrations are treated democratically, we anticipate the procedure to hold in both adiabatic and nonadiabatic regimes,
  since either $\delta n_{\sigma}$ or $\delta x_b$ can be made small, in contrast to the usual adiabatic mean-field treatment \cite{Galperin.05.NL}. 

%
\begin{figure}[tbh!]
 \centering
\includegraphics[width=0.8\columnwidth]{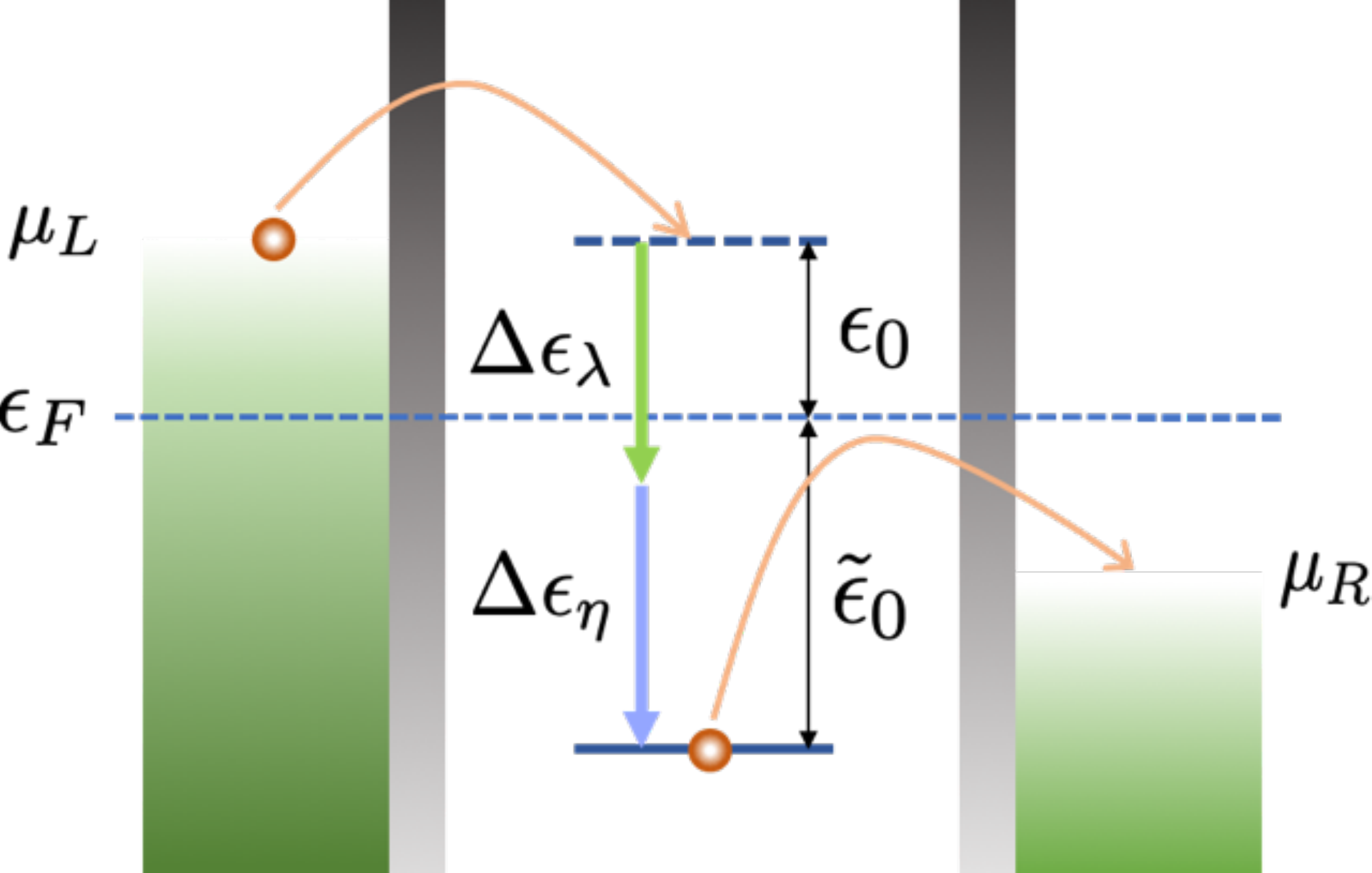} 
\caption{A schematic picture for NDR arising from QEVC. 
Metallic leads are drawn as continua with electrons filling up to the corresponding chemical potentials $\mu_{L,R}$. Single electronic levels are drawn as heavy horizontal dashed (solid) lines corresponding to the junction before (after) charging with level alignments $\epsilon_0$ ($\tilde{\epsilon}_0$) with respect to the Fermi energy $\epsilon_F$ (dashed line). 
Contributions to  level shift from the bilinear and quadratic electron-vibration couplings are marked by downward arrows with $\Delta\epsilon_{\lambda}$ and $\Delta\epsilon_{\eta}$, respectively. When the renormalized level moves outside the bias window, an NDR behavior occurs.} 
\label{fig:fig1}
\end{figure}

\begin{figure*}[tbh!]
 \centering
\includegraphics[width=2\columnwidth]{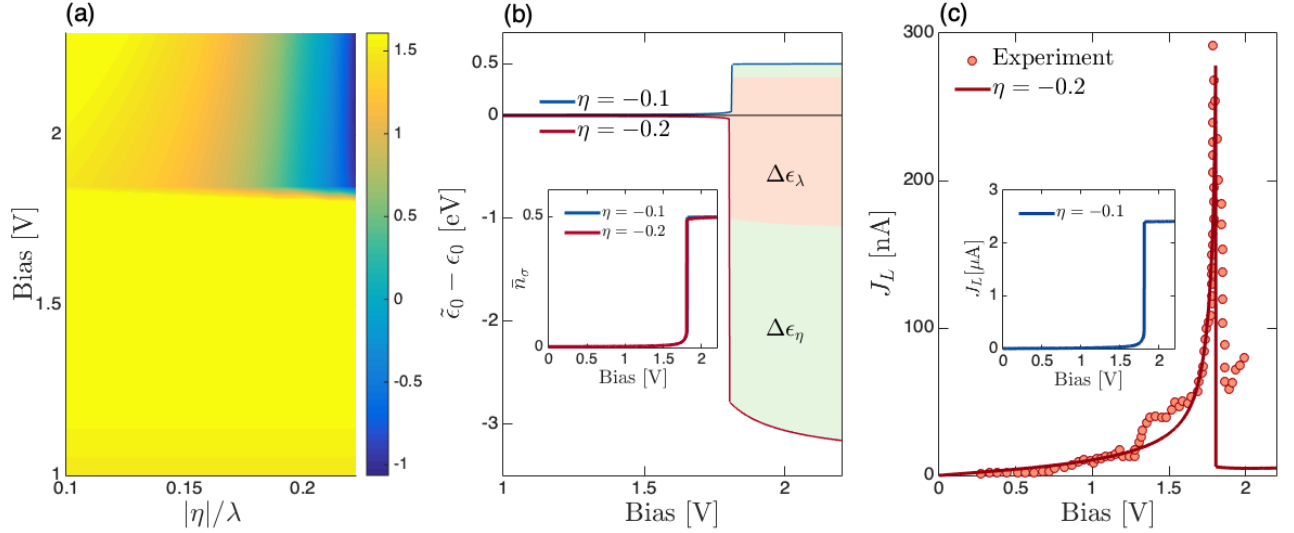} 
\caption{(a) Contour map for $\tilde{\epsilon}_0+V/2$ against voltage bias and the ratio $|\eta|/\lambda$. The region with $\tilde{\epsilon}_0+V/2<0$ indicates the occurrence of the predicted NDR effect as the renormalized electronic level moves outside the bias window. (b) Decomposition of voltage-dependent electronic level renormalization, $\tilde{\epsilon}_0-\epsilon_0$ for $\eta=-0.1$ (blue solid line) and $\eta=-0.2$ (red solid line) into its components  $\Delta\epsilon_{\lambda}$ (pink shaded region) and $\Delta\epsilon_{\eta}$ (green shaded region). Note that results for $\eta=-0.1$ are {\it reversed} in sign for contrast. 
The inset shows the average steady state charge occupation $\bar{n}_{\sigma=\uparrow,\downarrow}$ against applied voltage bias for $\eta=-0.1$ (blue solid line) and $-0.2$ (red solid line).
 (c) Calculated current-voltage characteristics for $\eta=-0.2$ (red solid line) shows an abrupt NDR behavior, which agrees well with  experimental measurements on OPE-NO$_2$ MJs (red circles) extracted from Fig. 2 (d) of Ref. \cite{Xiao.05.JACS}. The inset depicts the calculated current-voltage characteristics for $\eta=-0.1$, where the current saturates when entering the resonant transport regime. Other parameters are $\mu_L=-\mu_R=V/2$, $\epsilon_0=1$ eV, $\omega_b=0.138$ eV, $\lambda=0.9$, $\Gamma_L=\Gamma_R=5$ meV and $T=300$ K.} 
\label{fig:fig2}
\end{figure*}
%

The quadratic term $\eta\omega_b(b^{\dagger}+b)^2\sum_{\sigma}\bar{n}_{\sigma}$ can be eliminated by defining a dressed (Bogoliubov) mode: $a\equiv b\cosh r+b^{\dagger}\sinh r$, with the parameter $r$ determined from
 $e^r=\sqrt{\omega_a/\omega_b}$, and a renormalized vibrational frequency
\begin{equation}\label{eq:omega_a}
\omega_a=\omega_b\sqrt{1+4\eta \sum_{\sigma}\bar{n}_{\sigma}}.
\end{equation}
A negative $\eta$  leads to mode softening. The above form also leads to a constraint on the possible negative values of $\eta$ that we can adopt requiring $1+4\eta \sum_{\sigma}\bar{n}_{\sigma}>0$. Outside this region, there are two scenarios:
Bond breaking occurs, or higher order EVC terms, beyond the QEVC, stabilize the bond. These aspects are beyond
the scope of our modelling. 
Mode softerning based on Eq. (\ref{eq:omega_a}) is further discussed in \cite{SM.20.NULL}. 
Altogether, we arrive at an effective molecular Hamiltonian,
\bea\label{eq:hm2}
\tilde{H}_{M} &=& \tilde{\epsilon}_{0}\sum_{\sigma}n_{\sigma}+\omega_aa^{\dagger}a+e^{-r}\lambda \omega_b(a^{\dagger}+a)\sum_{\sigma}\bar{n}_{\sigma}.
\eea
Here, we defined a renormalized electronic energy (note that displacements of the original and dressed modes are equal, $\bar{x}_b=\bar{x}_a$),
\begin{equation}
\label{eq:epsilon_0_t}
\tilde{\epsilon}_{0}=\epsilon_0+\Delta \epsilon_{\lambda}+\Delta \epsilon_{\eta},
\end{equation}
with $\Delta\epsilon_{\lambda}\equiv\lambda\omega_b\sqrt{2\omega_b}\bar{x}_a$ and $\Delta \epsilon_{\eta}\equiv2\eta\omega_b^2\bar{x}_a^2$ denoting level shift due to LEVC and QEVC, respectively.  Eq. (\ref{eq:epsilon_0_t}) for the renormalized electronic energy is one of the key findings of this work. 
Notably, a polaron model \cite{Galperin.05.NL} is recovered when $\eta=0$.

We immediately notice that both contributions to the level renormalization are negative:  
$\lambda\bar{x}_a$ is always negative, see Eq. (\ref{eq:xa}) below. As well, $\eta$ is negative in general, see e.g. \cite{Kaasbjerg.13.PRB}.
While $\Delta \epsilon_{\lambda}$ alone is not large enough to induce an NDR under physical conditions as noted in Refs. \cite{Galperin.05.NL,Yeganeh.07.JACS,Fung.19.NL}, the situation changes once the contribution $\Delta \epsilon_{\eta}$ from the QEVC is taken into account. As we shall show below, a relatively small $\eta$ (compared to $\lambda$) can induce a significant level shift $\Delta\epsilon_{\eta}$ due to the square of $\bar{x}_a$ involved. Hence, the renormalized level $\tilde{\epsilon}_0$ can move outside the bias window upon charging as illustrated in Fig. \ref{fig:fig1}, resulting in an abrupt NDR behavior. We thus have one of the key findings of this work: QEVC and the associated mode softening can induce a sharp NDR effect in MJs. 

{\it Current-voltage characteristics.}--The efficacy of our theoretical treatment can be accessed by comparisons with experimentally observed current-voltage characteristics. Using our recently developed {\it nonperturbative} generalized input-output method, tailored for MJs \cite{Liu.20.PRB,Liu.20.PRBa},
and based on the effective molecular Hamiltonian Eq. (\ref{eq:hm2}), we get 
the steady state charge current out of the left lead as \cite{SM.20.NULL} 
\begin{equation}\label{eq:JL}
J_{L}~=~2\int\,\frac{d\epsilon}{2\pi}\frac{4\Gamma_L\Gamma_R}{\Gamma^2+(\epsilon-\tilde{\epsilon}_{0})^2}[n_F^L(\epsilon)-n_F^R(\epsilon)].
\end{equation}
Here the prefactor $2$ accounts for spin degeneracy, $\Gamma=\Gamma_L+\Gamma_R$ and $n_F^v(\epsilon)=\{\exp[(\epsilon-\mu_v)/T]+1\}^{-1}$ with $\mu_v$ the chemical potentials and $T$ the temperature denotes the Fermi-Dirac distribution function for the $v$ lead. 
Together with Eq. (\ref{eq:epsilon_0_t}) and steady state averages (see details in Ref. \cite{SM.20.NULL}),
\bea
\bar{x}_a &=& -\frac{\lambda\sqrt{2\omega_b}\sum_{\sigma}\bar{n}_{\sigma}}{(1+4\eta \sum_{\sigma}\bar{n}_{\sigma})\omega_b},\nonumber\\
\bar{n}_{\sigma} &=& 2\sum_v\Gamma_v\int\,\frac{d\epsilon}{2\pi}\frac{n_F^v(\epsilon)}{\Gamma^2+(\epsilon-\tilde{\epsilon}_{0})^2},
\label{eq:xa}
\eea
we approach the current-voltage characteristics in a self-consistent manner, by iterating the calculation of  $\bar{x}_a$ and
$\bar{n}_{\sigma}$ till convergence is reached \cite{SM.20.NULL}. 
Notably,  $\bar{n}_{\sigma}$ coincides with that obtained by the nonequilibrium Green's function method \cite{Galperin.05.NL} (noticing the adopted definition of $\Gamma_v$ is half of theirs). 

{\it Case study.}--We focus on the OPE-NO$_2$ single-molecule break junctions \cite{Xiao.05.JACS},
which display an abrupt NDR effect, as well as a diode behavior with the NDR feature showing only in one branch of the voltage bias.
A list organizing parameters, along with experimentally-relevant values (some from {\it ab initio} simulations) employed in our calculations
 are given in Table \ref{table1}. 
Elaborating: The value of $\epsilon_0$ for the LUMO (noting OPE-NO$_2$ is a LUMO-conducting molecule) is inferred from the HOMO (highest occupied molecular orbital)-LUMO gap \cite{Li.07.CMS} and the current-voltage characteristics \cite{Xiao.05.JACS}. $\omega_b$ takes the frequency of a ring mode, which participates in transport \cite{Selzer.05.NL}. $\lambda$ is set by the total reorganization energy, that is, $\lambda^2\omega_b\sim 0.11$ eV \cite{Yeganeh.07.JACS} (notably, only the total reorganization energy matters in charge transport as $\tilde{\epsilon}_0$ depends on the combination $\lambda^2\omega_b$). 
As for the range of $\eta$, while we do not have {\it ab initio} data for the OPE-NO$_2$ molecule, 
we set it by noting that (i) $\eta$ is bounded from below by $-0.25$, since we  enforce that $1+4\eta>0$ for charged molecules, with $\sum_{\sigma}\bar{n}_{\sigma}=1$ in the sequential tunneling regime  \cite{SM.20.NULL}. (ii)  Mode softening in OPE-NO$_2$ molecule is substantial \cite{Xiao.05.JACS}.
 (iii) From studies of similar molecules we learn that the ratio $|\eta|/\lambda$ ranges from 0 to 0.8  \cite{Kaasbjerg.13.PRB}. We emphasize that $\eta$ is the only freely-varying parameter in our calculations. 

\begin{table}[h]
\centering
\caption{Summary of parameters}
\begin{tabularx}{\columnwidth}{p  {1 cm}    l l  }
\hline
\hline
$\epsilon_0$  & Bare molecular electronic energy & $\sim$1 eV \cite{Li.07.CMS,Xiao.05.JACS}\\ 
$\Gamma_{L,R}$ & Hybridization energy to the metals & 0.1-10 meV \cite{Gehring.19.NRP}\\
$V$ & Voltage bias &  [-2.5 V, 2.5 V] \cite{Xiao.05.JACS}\\ 
$\omega_{b}$  & Frequency of primary mode & 0.138 eV \cite{Selzer.05.NL}\\
$\lambda$  & Bilinear electron-vibration coupling & 0.9 \cite{Yeganeh.07.JACS}\\ 
$T$ & Temperature of environments & 300 K \cite{Xiao.05.JACS}\\ 
$\eta$  & Quadratic electron-vibration coupling & -0.2-0 \\ 
\hline
\hline
\end{tabularx}
\label{table1}
\end{table}

To demonstrate the NDR mechanism, we first consider a scenario with a symmetric bias drop, that is, $|\mu_L|=|\mu_R|$.
 Characteristic simulation results are depicted in Fig. \ref{fig:fig2}. More details and comprehensive examples can be found in  \cite{SM.20.NULL}. 
 Since  $\tilde{\epsilon}_0<-|V|/2$  implies that the renormalized level shifts outside the bias window,  and the condition for an NDR is fulfilled, we focus on the behavior of $\tilde{\epsilon}_0+|V|/2$. A contour map for this measure is shown in Fig. \ref{fig:fig2} (a). As expected,  this indicator takes negative values in the parameter regime when (i) charging takes place with $\bar{n}_{\sigma}$ becoming large and (ii) $\eta$ is relatively large so as to ensure a significant contribution of $\Delta\epsilon_{\eta}$.

To gain more insights into level renormalization in the charging regime, 
we plot $\tilde{\epsilon}_0$ against the voltage bias for two representative values, $\eta=-0.1, -0.2$, in Fig. \ref{fig:fig2} (b). 
 Results for $\eta=-0.1$ are reversed in sign for clarity. 
 The contributions from $\Delta\epsilon_{\lambda}$ and $\Delta\epsilon_{\eta}$ to $\tilde{\epsilon}_0$ are further indicated. For $\eta=-0.1$,  level renormalization is mild such that one always find it within the bias window. The contribution from the QEVC $\Delta\epsilon_{\eta}$ is relatively small compared to $\Delta\epsilon_{\lambda}$. Decreasing $\eta$ to $-0.2$, we see a significant increase in the magnitude of $\Delta\epsilon_{\eta}$, resulting in a pronounced level renormalization, which shifts the level outside the bias window. Interestingly, although $\tilde{\epsilon}_0$ for $\eta=-0.1, -0.2$ depict distinct behaviors, we show in the inset that the average steady state charge occupation $\bar{n}_{\sigma}$ 
 against the voltage bias for the two values of $\eta$ almost coincides, thereby indicating that it is the QEVC, and not the plain charging effect, which plays the detrimental role in level renormalization and the resulting NDR effect, in accordance with experiments \cite{Xiao.05.JACS}. 

It is also worthwhile to mention that $\omega_a\sim\sqrt{1+4\eta}\omega_b\sim0.45\omega_b$ in the charged regime for $\eta=-0.2$. One may argue that this 
softening is quite large to be physical. However, we point out that here we only consider a single primary mode, hence the so-obtained renormalization should be regarded as an overall effect of many vibrational modes exhibiting frequency softening upon charging, which can be significant as showed by an {\it ab initio} simulation \cite{Kaasbjerg.13.PRB}. Nevertheless, the required $\eta$ value for the occurrence of the NDR is smaller if we consider scenarios with an asymmetric bias drop \cite{SM.20.NULL}. 

The calculated current-voltage characteristics for $\eta=-0.2$ is depicted in Fig. \ref{fig:fig2} (c). We obtain a very good agreement with experimental measurements \cite{Xiao.05.JACS}, thereby clearly demonstrating the viability of the proposed mechanism. Remarkably, we obtain this result by just including the QEVC on top of conventional bilinear modelings, and without fine-tunning $\eta$.
In fact, we have verified that the sharp NDR feature is rather robust for different $\eta$ values that fulfill the NDR condition, see also \cite{SM.20.NULL}.
In the absence of NDR, the charge current instead saturates in the resonant transport regime as the inset for $\eta=-0.1$ shows. Moreover, from the comparison between the insets of Fig. \ref{fig:fig2} (b)(c), we see that $J_L\propto\bar{n}_{\sigma}$ for $\eta=-0.1$, thereby indicating that a general relation uncovered for scenarios with only LEVCs \cite{Bijay.17.JCP} remains quantitatively valid in systems with relatively weak QEVCs.

We highlight the nature of the abrupt NDR. 
At this point, the energy level shifts deep below the bias window. 
If level occupation is not constrained to $\sum \bar n_{\sigma}\leq 1$, it can sharply rise approaching double occupancy, and
one gets $\omega_a^2<0$ \cite{SM.20.NULL}, which corresponds to bond dissociation. 
Indeed, as was observed e.g. in \cite{Xiao.05.JACS}, in many cases the sharp rise of the current was followed by the current dropping to zero, interpreted as voltage-induced molecular breakdown. 
Alternatively, if electron occupation on the dot is constrained due to Coulomb interaction, while bond softening is significant, it does not rupture, and the charge current after the drop remains finite albeit small corresponding to a valid NDR effect \cite{Xiao.05.JACS}. 
While we have checked that the abrupt feature of the IV curve is the same in both scenarios, to assess the stability of the junction beyond the peak, a more elaborate model is required.

\begin{figure}[tbh!]
 \centering
\includegraphics[width=0.95\columnwidth]{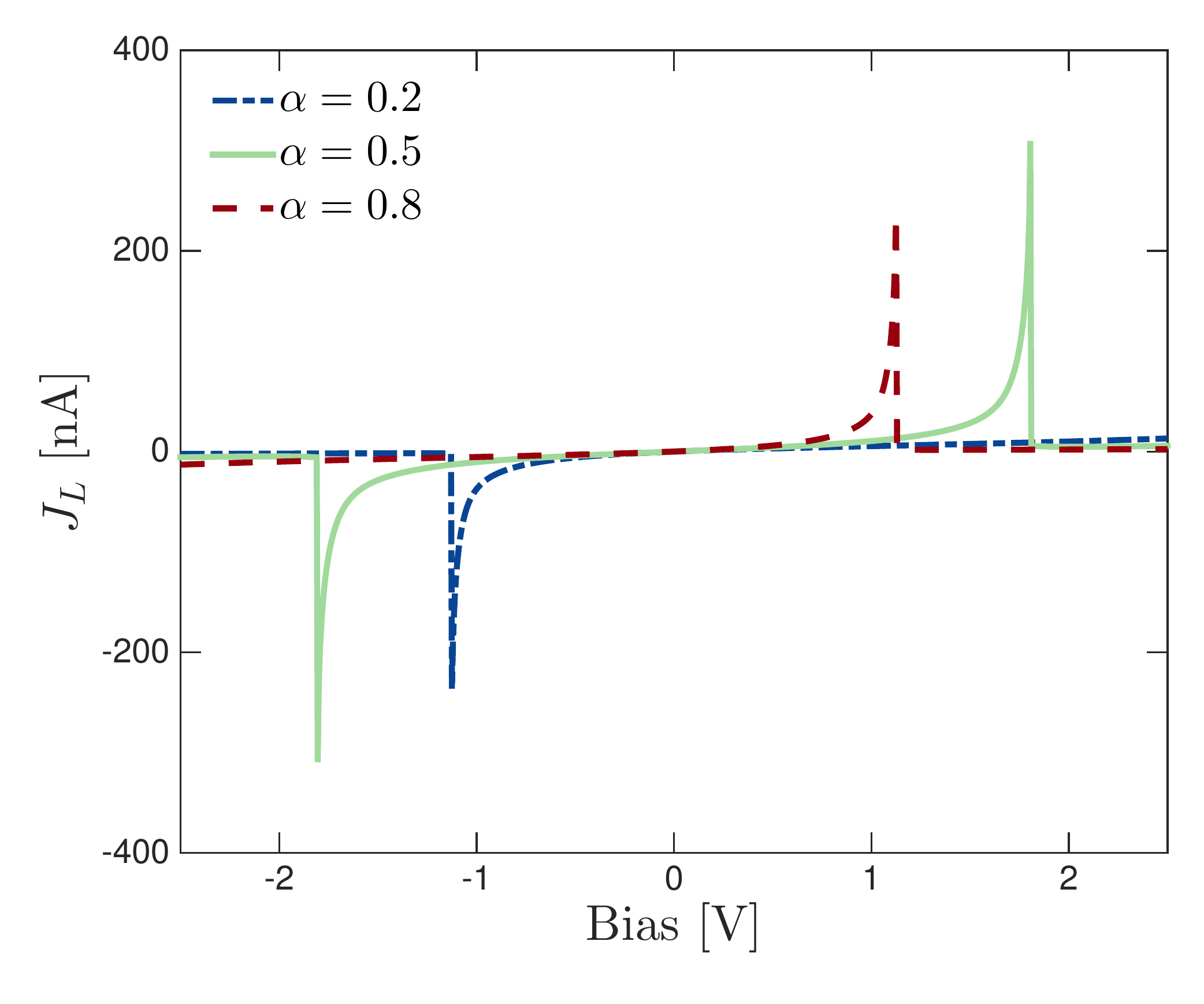} 
\caption{NDR and a diode effect calculated under bias drop $\mu_L=\alpha V$ and $\mu_R=-(1-\alpha)V$ for $\alpha=0.2$ (blue dash-dotted line), $\alpha=0.5$ (green solid line) and $\alpha=0.8$ (red dashed line). We use $\eta=-0.2$, while other parameters are the same as in Fig. \ref{fig:fig2}.} 
\label{fig:fig3}
\end{figure}
To mimic the situation where the asymmetric location of NO$_2$ group induces an asymmetric distribution of electric field across the molecule \cite{Xiao.05.JACS,Li.07.CMS}, we allow asymmetric bias drop on the junction. This effect is captured by introducing a phenomenological parameter $\alpha\in[0,1]$ such that the chemical potentials read $\mu_L=\alpha V$ and $\mu_R=-(1-\alpha) V$. A symmetric bias drop is recovered when $\alpha=0.5$. By recalling that the NDR occurs whenever the condition $\tilde{\epsilon}_0<\min(\mu_L,\mu_R)$ is fulfilled, we infer the following: 
For $\alpha\in[0,0.5)$, we have $|\mu_L|<|\mu_R|$ and hence the NDR effects appears in the negative ($\mu_R>\mu_L$) bias regime. In contrast, 
for $\alpha\in(0.5,1]$, the reverse holds. In both scenarios with $\alpha\neq0.5$, we also expect that the required voltage bias and $\eta$ value for the occurrence of NDR would become relatively smaller as compared to the symmetric case displayed in Fig. \ref{fig:fig2}. These predictions are confirmed by simulations depicted in Fig. \ref{fig:fig3}, see also \cite{SM.20.NULL}. 
Interestingly, we see from Fig. \ref{fig:fig3} that under a significant asymmetry the NDR peak can only be observed in the negative (positive) bias regime for $\alpha<0.5$ ($\alpha>0.5$) within the considered bias range; beyond this range, MJs typically break down. 
This one-side feature, a pronounced diode effect, is qualitatively consistent with experimental measurements \cite{Xiao.05.JACS}.

{\it Conclusions.}--We demonstrated that nonlinear IV characteristics in MJs are enacted by bond softening effects, particularly so-far largely overlooked QEVC.
We adopted a minimal model with both LEVC and QEVC and demonstrated an abrupt NDR behavior. We suggested that this mechanism could explain observed NDR behavior in OPE-NO$_2$ MJs.
 Future theoretical and computational studies with higher-order EVCs could clarify on the junction's stability beyond the NDR feature.
The operator-splitting procedure [Eq. (\ref{eq:splitting})] put forward here, which is valid in both adiabatic and nonadiabatic regimes, should be applicable for other systems exhibiting a time-scale separation. 
Our hope is that the demonstration presented here will inspire further theoretical, computational, and experimental studies on the critical role of high-order EVCs in the function of molecular devices.

The authors thank Zhenfei Liu for helpful discussions and 
acknowledge support from the Natural Sciences and Engineering Research Council (NSERC) 
of Canada Discovery Grant and the Canada Research Chairs Program.

%

\clearpage
\renewcommand{\thesection}{\Roman{section}} 
\renewcommand{\thesubsection}{\Alph{subsection}}
\renewcommand{\theequation}{S\arabic{equation}}
\renewcommand{\thefigure}{S\arabic{figure}}
\renewcommand{\thetable}{S\arabic{table}}
\setcounter{equation}{0}  
\setcounter{figure}{0}

\begin{widetext}
{\Large{\bf Supplemental material:} Sharp negative differential resistance from vibrational mode softening in molecular junctions}
\\
\\

In this supplementary material we present a thorough theoretical derivation of the steady state charge current expression as well as steady state averages used in the main text by resorting to a generalized input-output method tailored for molecular junctions (MJs) \cite{Liu.20.PRB,Liu.20.PRBa}. We also describe the self-consistent iterative scheme used in calculations, as well as numerical details for studying nitro-substituted oligo(phenylene ethynylene) (OPE-NO$_2$) single-molecule break junctions \cite{Xiao.05.JACS}.

\section{I. Generalized input-output method: Charge current and dynamical averages}
\label{a:1}
The total molecular junction Hamiltonian is $H=H_M+H_E$, with
\bea
H_M&=&\left[\epsilon_0+\lambda \omega_b(b^{\dagger}+b)+\eta\omega_b(b^{\dagger}+b)^2\right]\sum_{\sigma}n_{\sigma}+\omega_bb^{\dagger}b,
\nonumber\\
H_E&=&\sum_{kv\sigma}\epsilon_{kv}c_{kv\sigma}^{\dagger}c_{kv\sigma}+\sum_{kv\sigma}t_{kv}(c_{kv\sigma}^{\dagger}d_{\sigma}+d^{\dagger}_{\sigma}c_{kv\sigma}). 
\eea
The energies and coupling terms are defined in the main text. After applying the operator-splitting scheme $n_{\sigma}=\bar{n}_{\sigma}+\delta n_{\sigma}$ and $b=\bar{b}+\delta b$ and neglecting products between fluctuations as described in the main text, we treat the electron-vibration couplings in a mean-field manner. The quadratic term $\eta\omega_b(b^{\dagger}+b)^2 \sum_{\sigma}\bar n_{\sigma}$ is eliminated once we transform
to the  Bogoliubov mode, $a\equiv b\cosh r+b^{\dagger}\sinh r$, with the squeezing parameter  $e^r=\sqrt{\omega_a/\omega_b}$, and a renormalized vibrational frequency
$\omega_a=\omega_b\sqrt{1+4\eta \sum_{\sigma}\bar{n}_{\sigma}}$.
In terms of the Bogoliubov mode, the transformed Hamiltonian is $\tilde H=\tilde H_M+H_E$, where
the effective molecular Hamiltonian $\tilde{H}_M$ is
 \begin{equation}\label{eq:hm3}
\tilde{H}_{M}~=~\tilde{\epsilon}_{0}\sum_{\sigma}n_{\sigma}+\frac{p_a^2}{2}+\frac{1}{2}\omega_a^2x_a^2+\sqrt{2\omega_b}\lambda\omega_bx_a\sum_{\sigma}\bar{n}_{\sigma}.
\end{equation}
Here, we introduced the mass-weighted momentum operator $p_a=i\sqrt{\frac{\omega_a}{2}}(a^{\dagger}-a)$ and coordinate operator $x_a=\sqrt{\frac{1}{2\omega_a}}(a^{\dagger}+a)$ for the dressed primary vibrational mode. 
The metal Hamiltonian as well as the metal-molecule coupling term in $H_E$ remain the same.

In the input-output framework, we define input (incoming) fields from the reservoirs \cite{Liu.20.PRB,Liu.20.PRBa}.
To define input fields from metallic leads, we write down Heisenberg equations of motion (EOMs) for $c_{kv\sigma}$:
\bea\label{eq:eom_fc}
\dot{c}_{kv\sigma} &=& -i\epsilon_{kv}c_{kv\sigma}-it_{kv}d_{\sigma},
\eea
where we have introduced the notation $\dot{A}\equiv dA/dt$. 
Using the formal solutions for the EOM above, we get
\bea\label{eq:sum_env}
\sum_kt_{kv}c_{kv\sigma}(t) &=& \sqrt{2\pi}d_{in}^{v\sigma}(t)-i\sum_kt_{kv}^2\int_{t_0}^td\tau e^{-i\epsilon_{kv}(t-\tau)}d_{\sigma}(\tau).
\eea
Here, $t_0$ is the initial time at which dynamical evolution begins, and we have defined the input fields from the two electrodes as
\bea
d_{in}^{v\sigma}(t) &\equiv& \frac{1}{\sqrt{2\pi}}\sum_kt_{kv}e^{-i\epsilon_{kv}(t-t_0)}c_{kv\sigma}(t_0),
\eea
with the following correlation functions \cite{Liu.20.PRB}
\begin{eqnarray}
\langle d_{in}^{v\sigma,\dagger}(t')d_{in}^{v'\sigma'}(t)\rangle &=& \delta_{vv'}\delta_{\sigma\sigma'}\Gamma_v\int \frac{d\epsilon}{2\pi^2}e^{-i\epsilon(t-t')}n_F^v(\epsilon),\nonumber\\
\langle d_{in}^{v\sigma}(t)d_{in}^{v'\sigma',\dagger}(t')\rangle &=& \delta_{vv'}\delta_{\sigma\sigma'}\Gamma_v\int \frac{d\epsilon}{2\pi^2}e^{-i\epsilon(t-t')}\left[1-n_F^v(\epsilon)\right].
\end{eqnarray}
Here, $n_F^v(\epsilon)=\{\exp[(\epsilon-\mu_v)/T]+1\}^{-1}$ is the Fermi-Dirac distribution function with $\mu_v$ the chemical potential and $T$ the temperature.

To proceed, we consider the wideband limit \cite{Wingreen.89.PRB} for the metallic leads without compromising the value of molecule-lead hybridization energy $\Gamma_v$, which can be large. We then simplify Eq. (\ref{eq:sum_env}) as
\bea\label{eq:finals_1}
\sum_kt_{kv}c_{kv\sigma}(t) &=& \sqrt{2\pi}d_{in}^{v\sigma}(t)-i\Gamma_vd_{\sigma}(t).
\eea
The above relation is exact in the wideband limit \cite{Liu.20.PRB}.
We now consider the Heisenberg EOM for an arbitrary molecular (electrons and primary mode) operator $\mathcal{O}$,
\bea
\do{\mathcal{O}} &=& i[\tilde{H}_{M},\mathcal{O}]-i\sum_{kv\sigma}t_{kv}\left([\mathcal{O},c_{kv\sigma}^{\dagger}d_{\sigma}]+[\mathcal{O},d_{\sigma}^{\dagger}c_{kv\sigma}]\right).
\eea
As the molecular system contains both fermionic and bosonic operators, we should treat them separately. 
To this end, we redefine quantum commutator and anti-commutator as 
$[A,B]_-\equiv[A,B]$ and $[A,B]_+\equiv\{A,B\}$, respectively.
The EOM for $\mathcal{O}$ can be expressed as
\begin{eqnarray}
\dot{\mathcal{O}} &=& i[\tilde{H}_{M},\mathcal{O}]_{-}-i\sum_{kv\sigma}t_{kv}\left(\mp c_{kv\sigma}^{\dagger}[\mathcal{O},d_{\sigma}]_{\pm}+[\mathcal{O},d_{\sigma}^{\dagger}]_{\pm}c_{kv\sigma}\right).
\end{eqnarray}
Here, the top sign applies if $\mathcal{O}$ is a fermionic operator; the bottom sign applies if 
$\mathcal{O}$ is bosonic. 
Making use of Eq. (\ref{eq:finals_1}), we obtain a Heisenberg-Langevin equation (HLE)
\begin{equation}
\label{eq:eom_o}
\dot{\mathcal{O}}~=~ i[\tilde{H}_{M},\mathcal{O}]_{-}-i\sum_v\mathbb{L}_{\pm}^v,
\end{equation}
where 
\bea
\mathbb{L}_{\pm}^v&\equiv&\mp\left(i\Gamma_v d_{\sigma}^{\dagger}+\sqrt{2\pi}d_{in}^{v\sigma,\dagger}\right)[\mathcal{O},d_{\sigma}]_{\pm}
+[\mathcal{O},d_{\sigma}^{\dagger}]_{\pm}\left(-i\Gamma_vd_{\sigma}+\sqrt{2\pi}d_{in}^{v\sigma}\right).
\eea
The main observable of interest in the steady state limit is the total charge current across the MJ.
Introducing the charge occupation number operator of the left lead (source), 
$n_L\equiv\sum_{k\sigma}c_{kL\sigma}^{\dagger}c_{kL\sigma}$, the charge current out of the $L$ metal is formally given by
\bea
J_L=-\frac{d}{dt}\langle n_L \rangle=i\sum_{k\sigma}t_{kL}\langle(c_{kL\sigma}^{\dagger}d_{\sigma}-d_{\sigma}^{\dagger}c_{kL\sigma})\rangle\equiv\sum_{\sigma}J_{L\sigma},
\eea 
with the average performed over a factorized initial state of the composite system.
In the language of the input field, using Eq. (\ref{eq:finals_1}), we get
\begin{equation}\label{eq:js}
J_{L\sigma}~=~2\left(\sqrt{2\pi}\mathrm{Im}\langle d_{\sigma}^{\dagger}d_{in}^{L\sigma}\rangle-\Gamma_L\bar{n}_{\sigma}\right).
\end{equation}
Here, ``Im" refers to an imaginary part. We emphasize that the above working expression is formally exact in the wide-band limit \cite{Liu.20.PRB}.
By using Eq. (\ref{eq:eom_o}), we find
\bea
\dot{d}_{\sigma}(t) &=& -(\Gamma+i\tilde{\epsilon}_{0})d_{\sigma}(t)-i\sqrt{2\pi}\sum_vd_{in}^{v\sigma}(t),\label{eq:eom_d}\\
\ddot{x}_a(t) &=& -\omega_a^2x_a(t)-\lambda\sqrt{2\omega_b}\omega_b\sum_{\sigma}\bar{n}_{\sigma}.
\label{eq:eom_x}
\eea
Here, we have defined $\Gamma=\sum_v\Gamma_v$ and $\ddot{A}=d^2A/dt^2$. The renormalized electronic energy reads
\begin{equation}\label{eq:ome_d}
\tilde{\epsilon}_0~=~\epsilon_0+\lambda\omega_b\sqrt{2\omega}\bar{x}_a+2\eta\omega_b^2\bar{x}_a^2.
\end{equation}
Introducing the {\it free} retarded Green's function of the primary mode $D_0^r(t)=-i\Theta(t)\langle[x_a(t),x_a(0)]\rangle$ with $\Theta(t)$ the Heaviside step function satisfying $(\frac{d^2}{dt^2}+\omega_a^2)D^r_0(t-t')=-\delta(t-t')$, we obtain a formal solution for Eq. (\ref{eq:eom_x}):
\begin{equation}
x_a(t)~=~x_{a,0}(t)+\int\,d\tau D^r_0(t-\tau)\lambda\sqrt{2\omega_b}\omega_b\sum_{\sigma}\bar{n}_{\sigma},
\end{equation}
where $x_{a,0}(t)=x_a(t_0)\cos\omega_at+\frac{p_a(t_0)}{\omega_a}\sin\omega_at$ denotes the free evolution of the primary mode. The steady state average displacement is then given by
\bea\label{eq:bar_x}
\bar{x}_a &=& \lambda\sqrt{2\omega_b}\omega_b\sum_{\sigma}\bar{n}_{\sigma}\tilde{D}^r_0[\omega=0]~=~-\frac{1}{\omega_a^2}\lambda\sqrt{2\omega_b}\omega_b\sum_{\sigma}\bar{n}_{\sigma}\nonumber\\
&=& -\frac{\lambda\sqrt{2\omega_b}\sum_{\sigma}\bar{n}_{\sigma}}{(1+4\eta \sum_{\sigma}\bar{n}_{\sigma})\omega_b}.
\eea
In arriving at the above steady state average, we have utilized the expression for free retarded Green's function (GF) of the primary mode $\tilde{D}_0^r[\omega]=\frac{1}{\omega^2-\omega_a^2}$ in the Fourier space. 

One can take into account the coupling of primary mode to a secondary thermal bath by replacing the free GF with a full one, $\tilde{D}^r[\omega]=\frac{1}{\omega^2-\omega_a^2+\kappa^2}$ where $\kappa$ denotes the damping coefficient of the primary mode to the thermal bath \cite{Galperin.05.NL}. However, we point out that
the secondary bath does not play a role in our NDR effect as the coupling between the primary mode and thermal bath is typically weak \cite{Galperin.05.NL}, which renders $\kappa/\omega_b\ll 1$. 

As for the electronic operator $d_{\sigma}$, we can directly write down the formal solution in the steady state limit
\begin{equation}
d_{\sigma}(t)~=~-i\sqrt{2\pi}\sum_v\int_{-\infty}^td\tau e^{-(\Gamma+i\tilde{\epsilon}_{0})(t-\tau)}d_{in}^{v\sigma}(\tau),
\end{equation}
from which we get the steady state charge occupation
\begin{equation}\label{eq:nd}
\bar{n}_{\sigma}~=~2\sum_v\Gamma_v\int\,\frac{d\epsilon}{2\pi}\frac{n_F^v(\epsilon)}{\Gamma^2+(\epsilon-\tilde{\epsilon}_{0})^2},
\end{equation}
and the steady state charge current out of the left lead
\begin{equation}\label{eq:jl}
J_{L}~=~2\int\,\frac{d\epsilon}{2\pi}\frac{4\Gamma_L\Gamma_R}{\Gamma^2+(\epsilon-\tilde{\epsilon}_{0})^2}[n_F^L(\epsilon)-n_F^R(\epsilon)]
\end{equation}
with a prefactor 2 accounting for the spin degeneracy.

\section{II. Self-consistent numerical scheme}
\label{a:2}
To evaluate the steady state charge current, we should solve the coupled equations (\ref{eq:ome_d}), (\ref{eq:bar_x}) and (\ref{eq:nd}) in a self-consistent manner. In the present study, we adopt the following step-by-step iterative scheme for a fixed voltage bias,
\begin{itemize}
\item Step 1:~Choose an initial occupation condition ``$n_d$-trial" for $\bar{n}_{\sigma}$ and obtain initial values for $\tilde{\epsilon}_{0}$ and $\bar{x}_a$ according to Eqs. (\ref{eq:ome_d}) and (\ref{eq:bar_x});
\item Step 2:~Evaluate $\bar{n}_{\sigma}$ based on Eq. (\ref{eq:nd});
\item Step 3:~Update values for $\tilde{\epsilon}_{0}$ and $\bar{x}_a$ based on results generated by the step 2;
\item Step 4:~Repeat steps 2-3 iteratively until we meet an error threshold $\xi_e$ for $\bar{n}_{\sigma}$;
\item Step 5:~Evaluate steady state charge current $J_{L}$ based on Eq. (\ref{eq:jl}) with the so-obtained $\tilde{\epsilon}_0$.
\end{itemize}
We set $\xi_e=10^{-4}$ throughout the study, which is small enough to get well-converged self-consistent solutions. Unless otherwise stated, we always adopt a forward bias sweep starting from zero to some finite values in calculations presented in the main text and below. 

In Figure \ref{sfig:fig1}-\ref{sfig:fig2}, we test this scheme with and without the quadratic coupling $\eta$ in
two cases previously studied in the literature, and show good agreement. These setups do not lead to the NDR effect.
In Fig. \ref{sfig:fig3} we include $\eta$ with parameters providing  the NDR observation (Fig. 2 in the main text). We
further discuss then how we handle simulations beyond the NDR region.

\subsection{A. Demonstration: Polaron model}
As a first demonstration of the above iterative scheme, we note that our theory reduces to the polaron model of Ref. \cite{Galperin.05.NL} when $\eta=0$. In Fig. \ref{sfig:fig1} we show that indeed we are able to reproduce their Fig. 4 (a) by using our self-consistent scheme.
%
\begin{figure}[tbh!]
 \centering
\includegraphics[width=0.5\columnwidth] {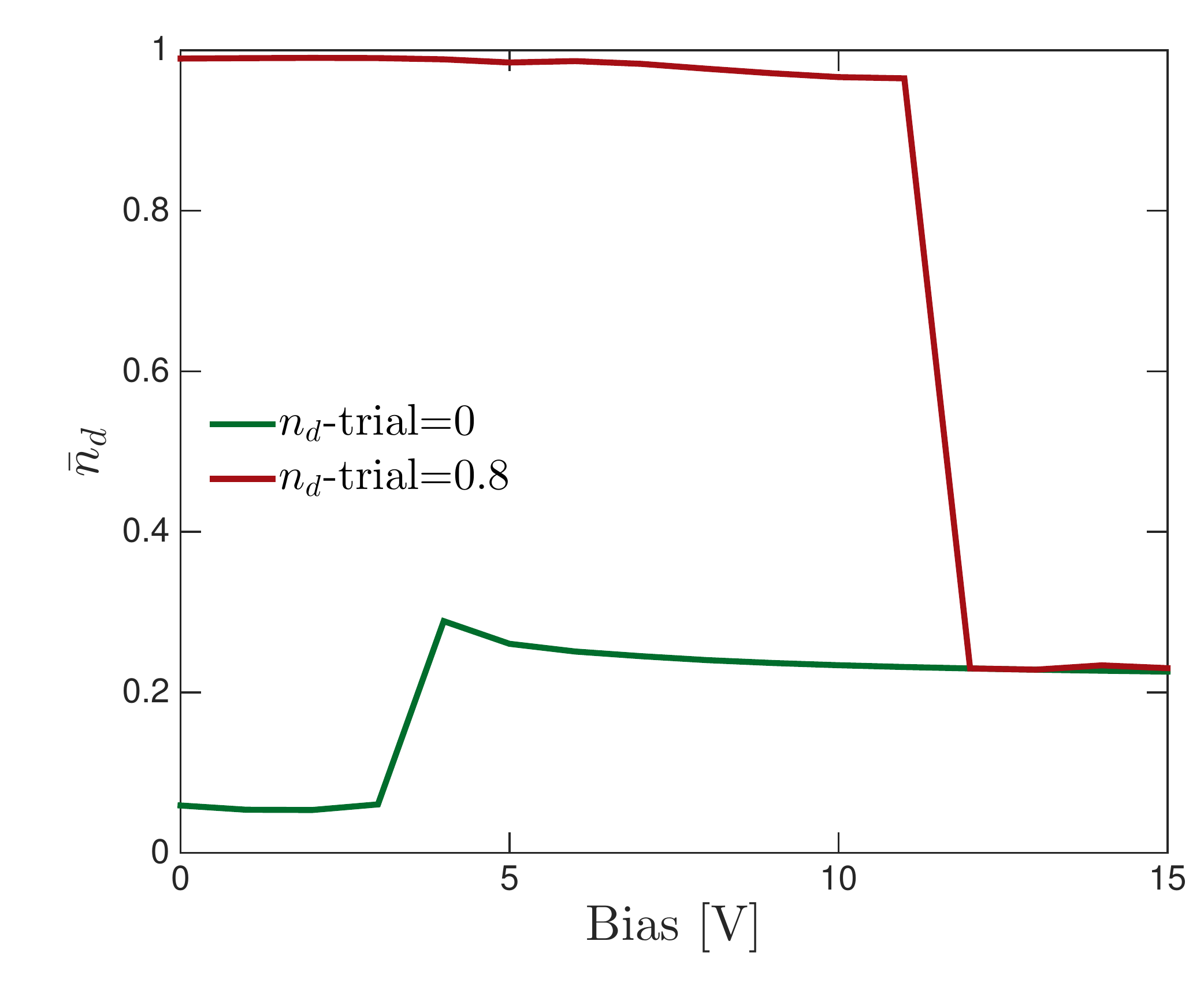} 
\caption{Steady state occupation $\bar{n}_d=\sum_{\sigma}\bar{n}_{\sigma}$ of a single level as a function of voltage bias $V$ for $n_d$-trial=0 (green line) and $n_d$-trial=0.8 (red line). Parameters are adopted from Ref. \cite{Galperin.05.NL}: $\mu_L=-\mu_R=V/2$, $\epsilon_0=3$ eV, $\omega_b=0.05$ eV, $\lambda\omega_b=0.4$ eV, $\Gamma_L=0.1$ eV, $\Gamma_R=0.35$ eV, $T=300$ K, and $\eta=0$. %
Here and below, the voltage is reported in volts.}  %
\label{sfig:fig1}
\end{figure}
As noted in Ref. \cite{Galperin.05.NL}, one can start from either an empty level situation or a filled level case. In our calculations. we take $n_d$-trial=0 and $n_d$-trial=0.8 (for the total occupation $\bar{n}_d\equiv\bar{n}_{\uparrow}+\bar{n}_{\downarrow}$) respectively. As can be seen from Fig. \ref{sfig:fig1}, $\bar{n}_d$ depicts distinct voltage bias dependence for different $n_d$-trial. However, this is not always the case for a polaron model as we will show in Fig. \ref{sfig:fig4} (a) with a different set of parameter values. There, results showed minimal sensitivity on the initial conditions.  


\subsection{B. Demonstration: OPV3 molecular junction}
As a second verification of our procedure, we apply the self-consistent iterative scheme together with $\omega_a=\omega_b\sqrt{1+4\eta \sum_{\sigma}\bar{n}_{\sigma}}$ to calculate the vibrational frequency renormalization for an oligo(3)-phenylenevinylene (OPV3) MJ.
This system was studied experimentally and theoretically in Refs. \cite{Ward.11.NN,Kaasbjerg.13.PRB}. A set of simulation results is depicted in Fig. \ref{sfig:fig2}. We verified that for the parameters adopted in Fig. \ref{sfig:fig2}, the fixed-point of the above self-consistent iterative scheme is unique.  
%
\begin{figure}[tbh!]
 \centering
\includegraphics[width=0.5\columnwidth] {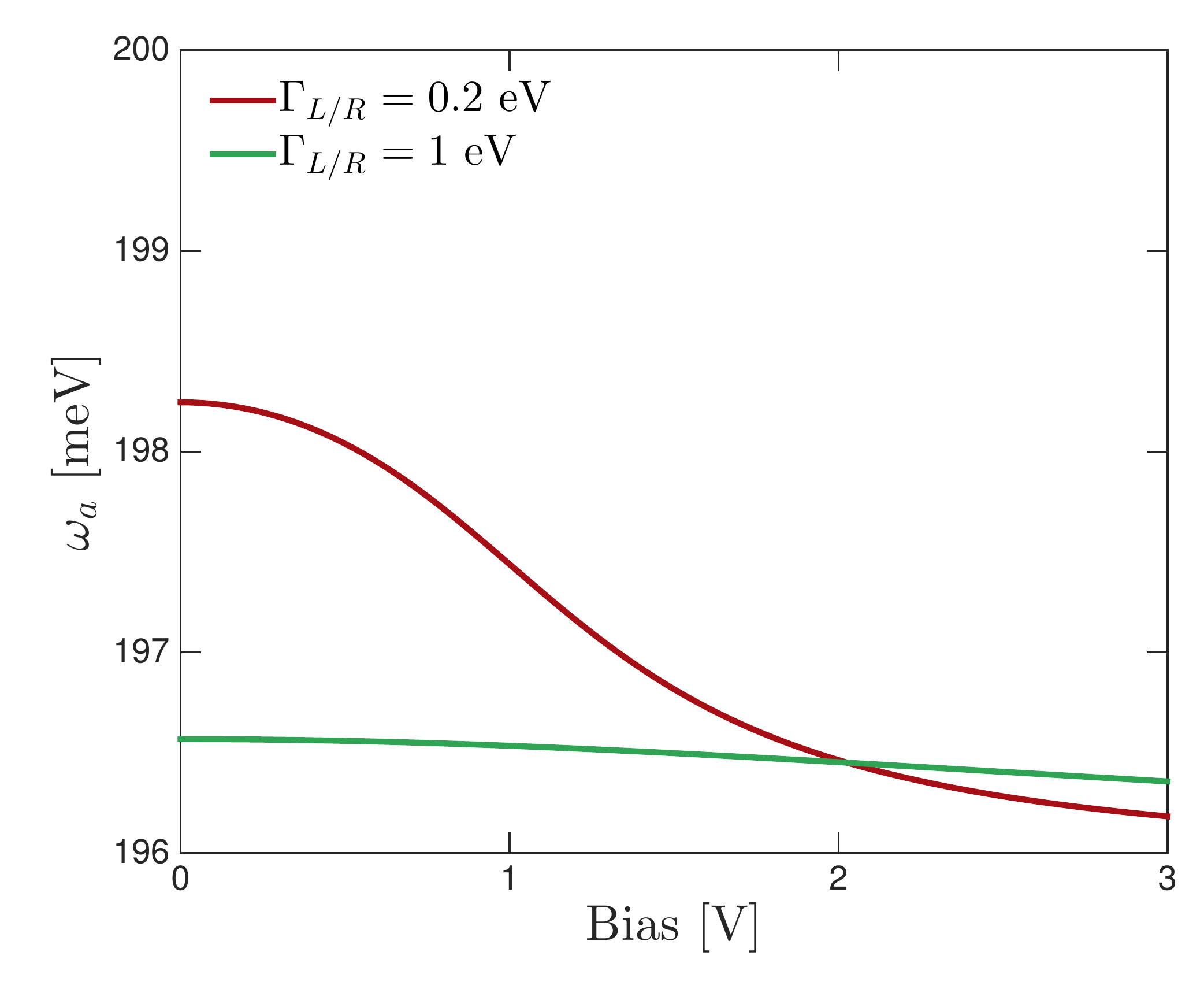} 
\caption{Renormalized vibrational frequency $\omega_a=\omega_b\sqrt{1+4\eta \sum_{\sigma}\bar{n}_{\sigma}}$ for the OPV3 MJ as a function of voltage bias $V$. Parameters are adopted from Ref. \cite{Kaasbjerg.13.PRB}: $\mu_L=-\mu_R=V/2$, $\epsilon_0=0.5$ eV, $\omega_b=200$ meV, $\lambda\omega_b=50$ meV, $\eta\omega_b=-2$ meV and $T=0$ K. 
} 
\label{sfig:fig2}
\end{figure}

As can be seen from the figure, our simple expression for $\omega_a$  closely reproduces  the frequency renormalization with voltage  $V$ and molecule-lead hybridization energy $\Gamma_{L,R}$ compared to a nonequilibrium Green's function method calculation \cite{Kaasbjerg.13.PRB}. However, due to the fact that our single-level model only includes the lowest unoccupied molecular orbital (LUMO) without involving the highest occupied molecular orbital (HOMO), the electron-hole pair excitation considered in Ref. \cite{Kaasbjerg.13.PRB} is beyond the scope of the present study. Hence, Fig. \ref{sfig:fig2} misses the contribution from Pauli blocking and it does not capture the follow-up hardening when $V>1.3$. 
Nevertheless, our simple model can quantitatively describe the current-induced vibrational frequency renormalization 
in LUMO-conducting MJs.
 
\section{III. Calculation details for OPE-NO$_2$ molecular junction}
\label{a:3}
\subsection{A. Properties of the self-consistent solutions}
Eq. (\ref{eq:ome_d}) is a second-order polynomial in $\bar{x}_a$. Hence, we may expect two solutions corresponding to the roots of that polynomial when $\eta$ becomes relatively large. However, because of the requirement that $\omega_a=\omega_b\sqrt{1+4\eta\sum_{\sigma}\bar{n}_{\sigma}}$ must be real, not all solutions are physical. In fact, we find that the system always supports only one physical steady state solution when using parameters for OPE-NO$_2$ MJ. A representative set of self-consistent solutions is shown in Fig. \ref{sfig:fig3} for $\eta=-0.1$ (a) and $-0.2$ (b).
%
\begin{figure}[tbh!]
 \centering
\includegraphics[width=0.7\columnwidth] {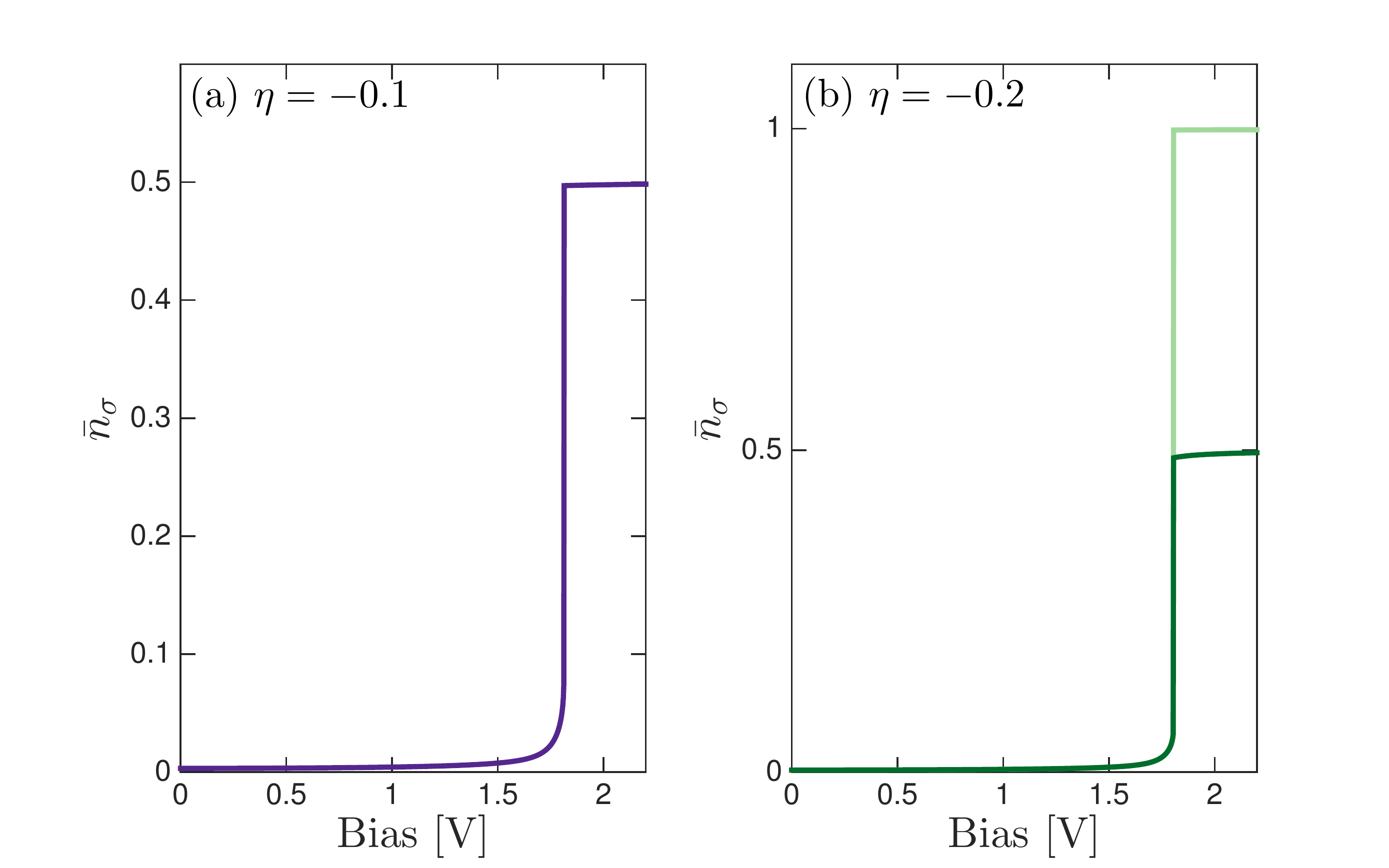} 
\caption{Average steady state occupation of the single level, $\bar{n}_{\sigma}$, as a function of voltage bias for (a) $\eta=-0.1$ and (b) $\eta=-0.2$.  In (b), there are two stable solutions, a physical case (dark line) and a non-physical solution (light), the latter leads to an imaginary frequency and is discarded.
Other parameters are $n_d$-trial=0, $\mu_L=-\mu_R=V/2$, $\epsilon_0=1$ eV, $\omega_b=0.138$ eV, $\lambda=0.9$, $\Gamma_L=\Gamma_R=5$ meV and $T=300$ K.} 
\label{sfig:fig3}
\end{figure}

When $\eta$ is relatively small, we see from Fig. \ref{sfig:fig3} (a) that the fixed-point of the self-consistent iterative scheme is unique and physical. In contrast, for $\eta=-0.2$ [Fig. \ref{sfig:fig3} (b)], we end up with two stable solutions in the charged regime once the contribution due to $\eta$ to $\tilde{\epsilon}_0$ [Eq. (\ref{eq:ome_d})] becomes significant. However, we notice that one of the solution has an average charge occupation per spin species of the order of 1, implying that $1+4\eta\sum_{\sigma}\bar{n}_{\sigma}<0$ ($\bar{n}_{\uparrow}=\bar{n}_{\downarrow}$ as will be shown in Fig. \ref{sfig:fig4}). While this value is acceptable for the self-consistent iteration scheme as $\bar{x}_a$ involves just $1+4\eta\sum_{\sigma}\bar{n}_{\sigma}$, it results in an imaginary vibrational frequency as $\omega_a$ depends on the square root $\sqrt{1+4\eta\sum_{\sigma}\bar{n}_{\sigma}}$. Hence the solution with $1+4\eta\sum_{\sigma}\bar{n}_{\sigma}<0$ is unphysical. Below and in the main text, we only depict the physical solution.

As we show in the main text (Fig. 2), once we increase the voltage, the level occupancy grows, 
$\omega_a=\omega_b\sqrt{1+4\eta \sum_{\sigma}\bar n_{\sigma}}$ reduces, the renormalized electronic energy drops, and the current quickly rises.
At a certain point, the square root term in the renormalized frequency becomes zero. How should we treat this point, and beyond?

Numerically, at this point the iterative scheme becomes unstable, with $\bar n_{\sigma}$ alternating between two values, close to 1/2, and close to 1. One possible scenario is that when $\omega_a$ approaches zero, bond breaking occurs.
In principle, the model then reduces to the noninteracting Anderson dot model, and one may continue simulations without the vibration.
Another scenario, which we adopt here, is to enforce the sequential tunneling limit and disallow the total level population to exceed one. 
This is reasonable as we assume small  hybridization $\Gamma$  and high temperature.
As such, once population begins to grow, we only permit solutions with $\sum_{\sigma}\bar n_{\sigma}\leq 1$.
Nevertheless, we emphasize that the current-voltage characteristics up to the NDR abrupt jump is identical whether or not we limit occupation on the level. 
Beyond that, at higher voltage, one needs to enrich our model: Stabilize the mode by including
electron-vibration couplings beyond the quadratic model, add several prominent modes that exchange energy, or take into account the coulomb repulsion energy to limit level occupation.

In Fig. \ref{sfig:fig4}, we present physical solutions ($\omega_a$ real-positive) for $\bar{n}_{\uparrow}$ (top column) and $\bar{n}_{\downarrow}$ (bottom column) against the initial occupation condition $n_d$-trial
 and voltage bias for three different values of $\eta$. Note that  we use here the 
 same initial condition at each voltage value, i.e. we restart the self consistent scheme with the trial occupation $n_d$-trial at each voltage point.
 First, we find that one always retains $\bar{n}_{\uparrow}=\bar{n}_{\downarrow}$, regardless of the value of $\eta$. This is expected as we consider a spin-degenerate scenario. Second, the {\it basic} trend of $\bar{n}_{\sigma}$ as a function of voltage bias is independent of the choice of $n_d$-trial, in a sharp contrast to Fig. \ref{sfig:fig1}. Hence in the calculations presented in the main text, we fix $n_d$-trial=0, which is reasonable as one should start from an empty state for a single level system with $\epsilon_0>0$ in the forward bias direction. 
 
 %
\begin{figure}[tbh!]
 \centering
\includegraphics[width=0.8\columnwidth] {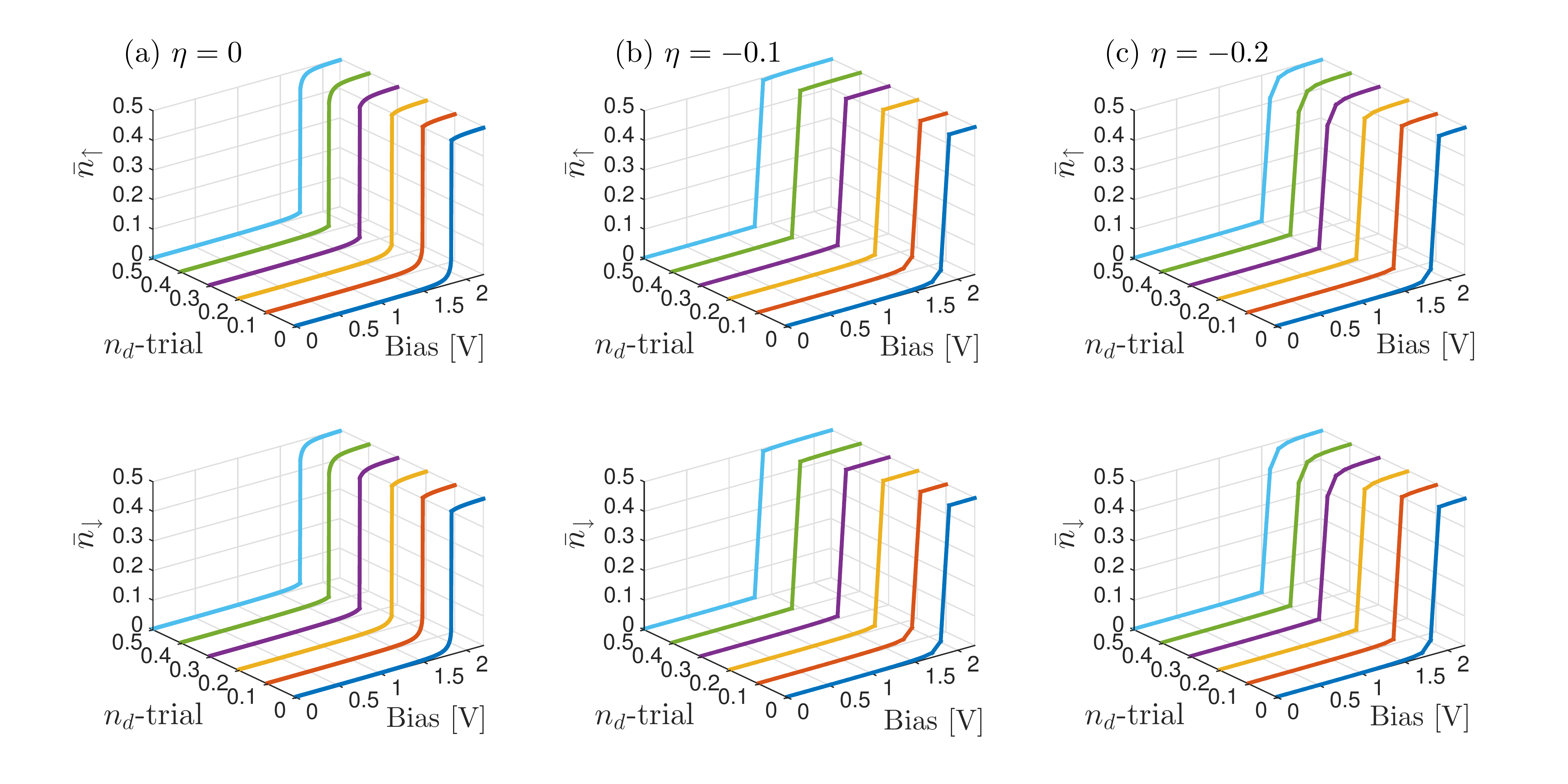} 
\caption{Average steady state occupation $\bar{n}_{\uparrow}$ (top column) and $\bar{n}_{\downarrow}$ (bottom column)  as a function of voltage bias and $n_d$-trial for (a) $\eta=0$ (left panel), (b) $\eta=-0.1$ (middle panel) and (c) $\eta=-0.2$ (right panel). Other parameters are the same as in Fig. \ref{sfig:fig3}.} 
\label{sfig:fig4}
\end{figure}
In Fig. \ref{sfig:fig5} (c), we further demonstrate that results remain the same even if we update  $n_d$-trial as we increase the voltage in the forward bias direction (compare to Fig. 2 in the main text). Nevertheless, we  point out that there is a slight difference between curves with $n_d$-trial=0 and 0.5; the latter requires a slightly lower voltage bias for the charging process to take place. This difference does not affect results shown in the main text and in the above as we considered a {\it forward} bias sweep starting from zero. However, if the forward protocol is further followed by a backward bias sweep, we should expect a nontrivial impact of the initial condition on current-voltage characteristics. This is because the backward protocol begins from a charged state, and it should be performed with the trial value $n_d$-trial=0.5. In fact, as we show in Fig. \ref{sfig:fig5}, a hysteresis behavior emerges in the current-voltage characteristics due to such a slight difference in the charging process.

\subsection{B. Hysteresis behavior} 
To check whether our results depend on the voltage bias sweep direction, we follow a procedure that is frequently adopted in experiments: The voltage bias is first swept from 0 V to 2.5 V (forward direction) and then from 2.5 V to 0 V (backward direction). 
%
\begin{figure}[tbh!]
 \centering
\includegraphics[width=0.8\columnwidth] {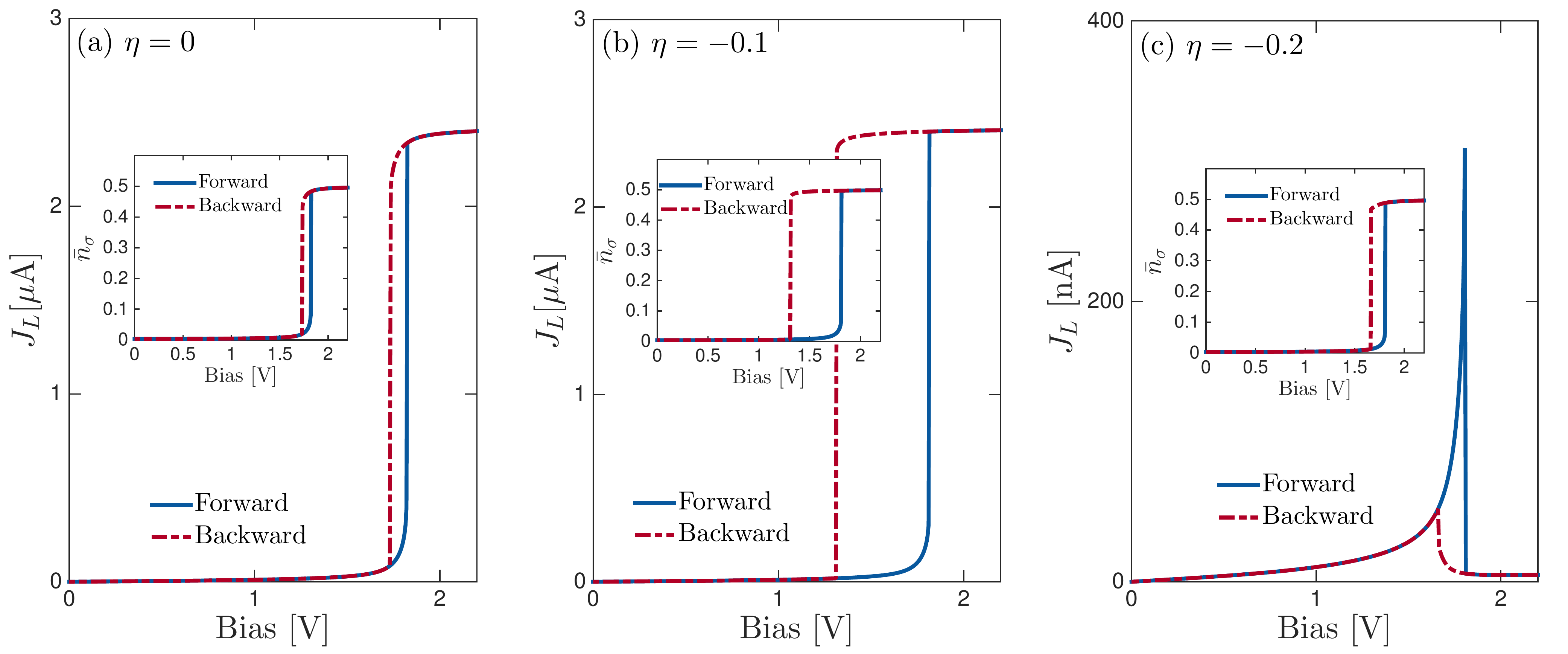}
\caption{Current-voltage characteristics under different bias sweep directions and with updated initial condition protocol for the 
level occupation. 
(a) $\eta=0$, (b) $\eta=-0.1$ and (c) $\eta=-0.2$. The voltage bias is first swept from 0 V to 2.5 V (forward, blue solid line), then from 2.5 V to 0 V (backward, red dash-dotted line). The insets show the average occupation number $\bar{n}_{\sigma}$ against the voltage bias with forward (blue solid line) and backward (red dash-dotted line) bias sweep. Other parameters are the same as in Fig. \ref{sfig:fig3}.} 
\label{sfig:fig5}
\end{figure}
To account for the fact that different sweep directions have different initial conditions as analyzed before, in simulations we update the value for $n_d$-trial as the applied voltage bias changes: for discretized voltage values $\{\cdots,V_{n}, V_{n+1},\cdots\}$ ordered in time, the converged physical solution for $\bar{n}_{\sigma}$ obtained for a voltage bias $V_n$ becomes the $n_d$-trial for the self-consistent iteration loop for $V_{n+1}$. For a small voltage increment $|V_{n+1}-V_n|$ (0.005 in simulations), this strategy faithfully captures the physical initial conditions.  As an illustration, we consider the scenario with symmetric bias drop and depict the corresponding current-voltage characteristics in Fig. \ref{sfig:fig5}. Clearly, from Fig. \ref{sfig:fig5}, we observe a hysteresis behavior of the current-voltage characteristics precisely due to the difference between charging processes in different bias directions in accordance with $\bar{n}_{\sigma}$ from Fig. \ref{sfig:fig4}. However, we have checked that if we fix $n_d$-trial (no matter which value we adopt, 0 or 0.5) during the whole voltage bias evolution, the hysteresis behavior disappears. Nevertheless, we see from Fig. \ref{sfig:fig5} (c) that the current-voltage characteristics obtained under the forward bias sweep while updating $n_d$-trial remains the same to that obtained by fixing $n_d$-trial=0, shown in Fig. 2 (c) of the main text. Hence, if one just considers the forward bias sweep, a fixed $n_d$-trial is a convenient choice.

\subsection{C. Temperature dependence of current-voltage characteristics}
We  illustrate  the temperature dependence of the current-voltage characteristics under symmetric bias drop in Fig. \ref{sfig:fig6}. From the inset, we find that the peak voltage is a monotonic decreasing function of temperature, in agreement with experimental observations for OPE-based self-assembled monolayers \cite{Chen.S.99}. Such a temperature dependence could enable further experimental verifications of our mechanism in the context of single MJ.
%
\begin{figure}[tbh!]
 \centering
\includegraphics[width=0.52\columnwidth] {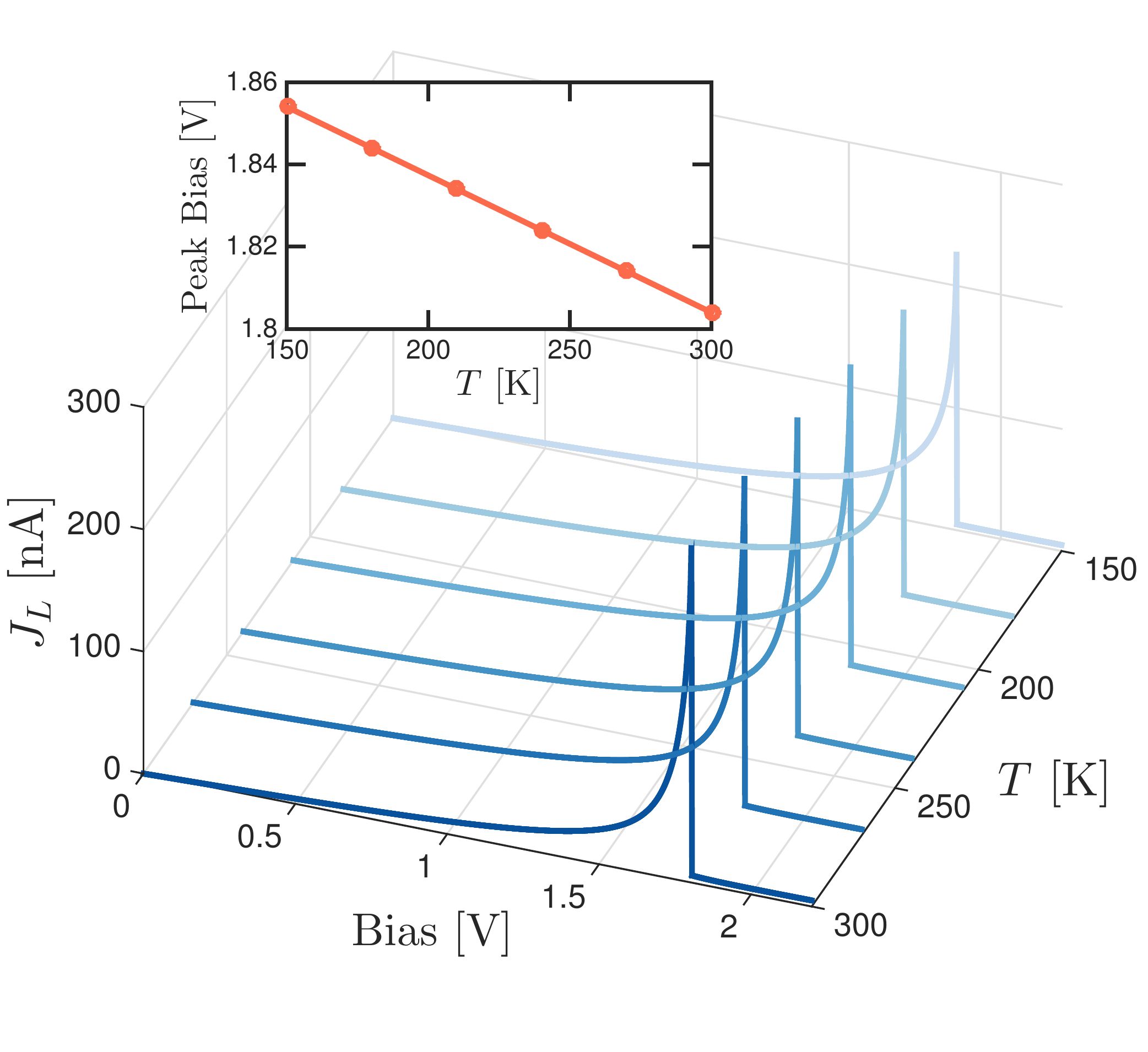} 
\caption{Steady state charge current $J_L$ as a function of voltage bias and temperature $T$. The inset displays the temperature dependence of the peak voltage. We take $\eta=-0.2$, other parameters are the same as in Fig. \ref{sfig:fig3}.} 
\label{sfig:fig6}
\end{figure}

\subsection{D. NDR effect for asymmetric bias drop}
An asymmetric bias drop is implemented using $\mu_L=\alpha V, \mu_R=-(1-\alpha)V$.
To supplement results shown in the main text, we depict in Fig. \ref{sfig:fig7} a detailed contour map for the indicators $\tilde{\epsilon}_0+|\mu_R|$ (used when $V>0$) and $\tilde{\epsilon}_0+|\mu_L|$  (once $V<0$). The parameter regime where these indicators become negative identify the occurrence of an abrupt NDR.
%
\begin{figure}[tbh!]
 \centering
\includegraphics[width=0.75\columnwidth] {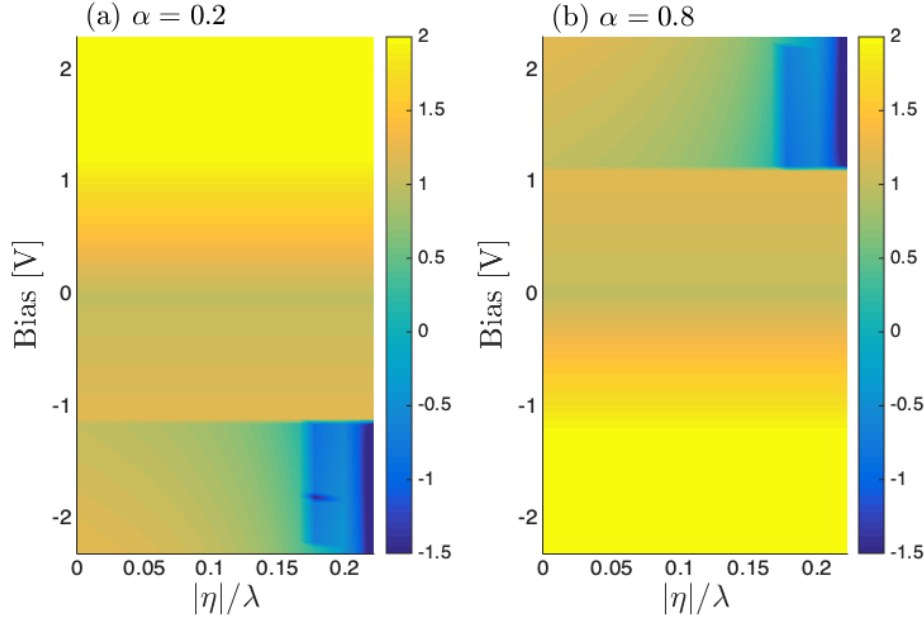} 
\caption{Contour map for the indicators $\tilde{\epsilon}_0+(1-\alpha)V$ ($V>0$) and
$\tilde{\epsilon}_0-\alpha V$ ($V<0$) as a function of voltage bias and the ratio $|\eta|/\lambda$ for (a) $\alpha=0.2$ and (b) $\alpha=0.8$. Other parameters are the same as in Fig. \ref{sfig:fig3}.} 
\label{sfig:fig7}
\end{figure}
We observe two important features from the figure: (i) NDR can only be observed in either negative or positive bias regime under large asymmetry depending on whether $\alpha<0.5$ or $\alpha>0.5$. (ii) The parameter regime where an NDR can occur becomes much broader than that with a symmetric bias drop in the sense that the NDR can be induced with smaller voltage bias and $\eta$ values. These observations are consistent with the general analysis given in the main text.

In Fig. \ref{sfig:fig8}, we further examine the sensitivity of the current-voltage characteristics showing an abrupt NDR behavior to the magnitude of $\eta$. We consider an asymmetric bias drop with $\alpha=0.8$ and three different $\eta$ values, which allow for the occurrence of the NDR.
%
\begin{figure}[tbh!]
 \centering
\includegraphics[width=0.5\columnwidth] {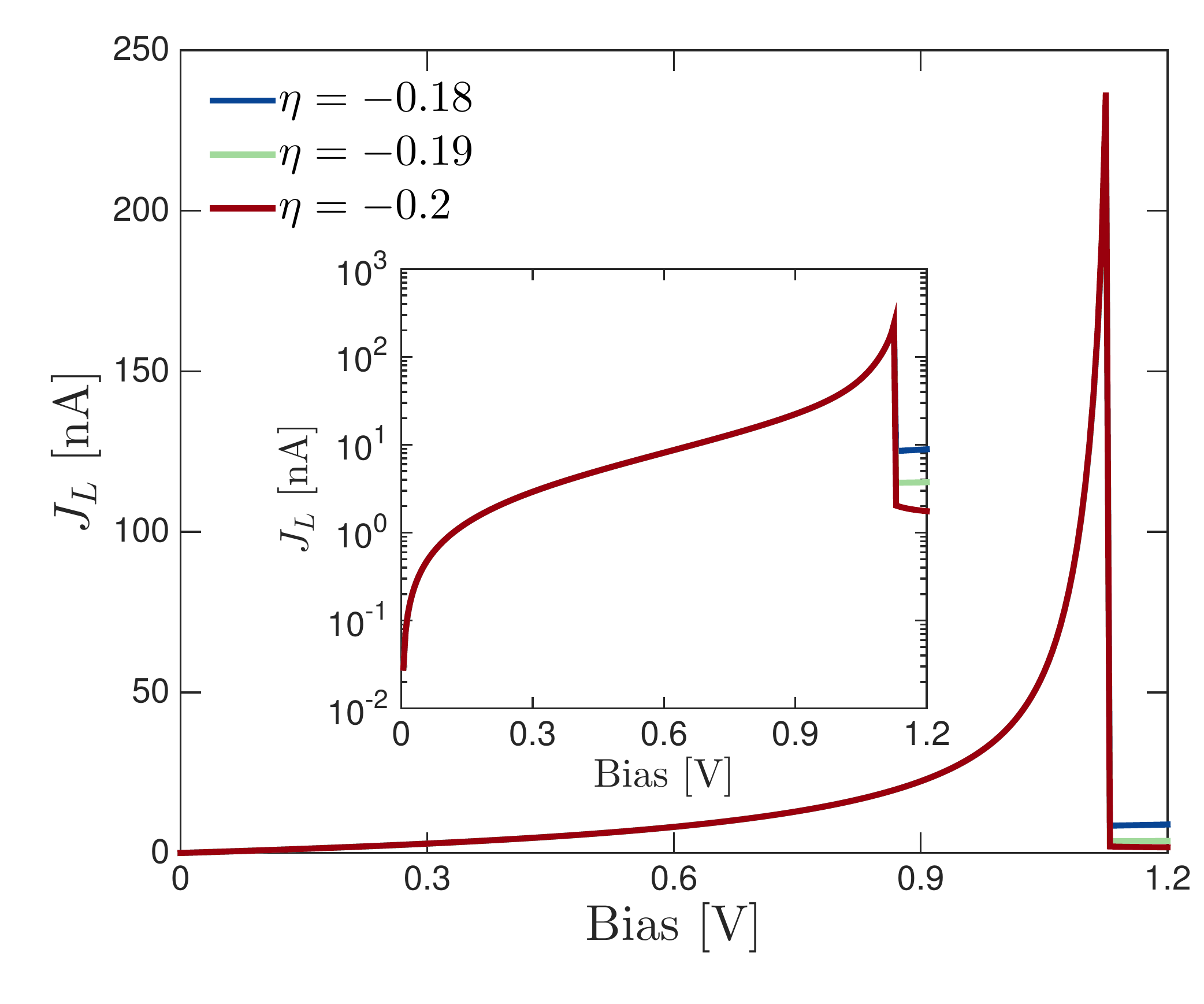} 
\caption{Current-voltage characteristics for $\eta=-0.18$ (blue solid line), $\eta=-0.19$ (green solid line) and $\eta=-0.2$ (red solid line). The inset depicts the same results on a logarithmic scale. We fix $\alpha=0.8$; other parameters are the same as in Fig. \ref{sfig:fig3}.} 
\label{sfig:fig8}
\end{figure}
Apparently, varying $\eta$  only affects the peak-to-valley ratio as highlighted by the inset where results are depicted on a logarithmic scale. 
In contrast, the sharp character is quite robust against possible $\eta$ values. This is expected as the current enhancement right before the sharp drop occurs is determined by the `bare' energy level. The value of $\eta$ (for those cases that fulfill the NDR condition) mainly sets the magnitude of level renormalization, and hence the peak-to-valley ratio.

\end{widetext}

\end{document}